\documentclass[12pt]{article}
\begin{document}



\begin{center}

{\Large Analytic solution of Hubbell's}

\smallskip

{\Large Model of Local Community Dynamics}

\bigskip

\bigskip

{\large Alan J. McKane$^{1}$, David Alonso$^{2,3}$ and 
Ricard V. Sol\'e$^{2,4}$}

\bigskip

\bigskip

{\small $^{1}$Department of Theoretical Physics, 
University of Manchester, Manchester M13 9PL, UK \\ 
$^{2}$Complex Systems Lab, Universitat Pompeu Fabra, Dr Aiguader 80,
08003 Barcelona, Spain \\ 
$^{3}$Department of Ecology, Facultat de Biologia, Universitat de Barcelona, \\
Diagonal 645, 08045 Barcelona, Spain \\
$^{4}$Santa Fe Institute, 1399 Hyde Park Road, New Mexico 87501, USA}
\end{center}

\vspace{2cm}

\begin{abstract}
Recent theoretical approaches to community structure and dynamics reveal that 
many large-scale features of community structure (such as species-rank 
distributions and species-area relations) can be explained by a so-called 
neutral model. Using this approach, species are taken to be equivalent and
trophic relations are not taken into account explicitly. Here we provide 
a general analytic solution to the local community model of Hubbell's neutral 
theory of biodiversity by recasting it as an urn model i.e.\,a Markovian 
description of states and their transitions. Both stationary and 
time-dependent distributions are analysed. The stationary 
distribution --- also called the zero-sum multinomial --- is given in closed 
form. An approximate form for the time-dependence is obtained by using an 
expansion of the master equation. The temporal evolution of the approximate
distribution is shown to be a good representation for the true temporal 
evolution for a large range of parameter values. 
\end{abstract}
\vspace{2 cm}

\vspace{5 cm}
\begin{center}
{\em \large Submitted to Theoretical Population Biology}
\end{center}

\newpage

\section{Introduction}
\label{intro}
Understanding the global patterns of biodiversity and its dynamics at 
different time scales remains a great challenge for ecological science 
(Rosenzweig, 1995; Wilson, 2003). One of the key features that defines 
community structure is the relation between range and abundance. How the
community structure develops in time and how species are spatially 
distributed largely define the field of macroecology (Brown, 1995). In this 
context, an important step to unify biogeography and biodiversity has been 
achieved by Hubbell (Hubbell, 2001; Bell, 2001) through the formulation of a 
neutral theory.

The mathematical framework employed by Hubbell allows for speciation processes 
to be integrated with the MacArthur-Wilson theory of island biogeography. 
In this way, the neutral theory predicts some universal features that can be 
tested by direct analysis of species-abundance distributions and other 
large-scale measurements. In Hubbell's theory, two key quantities largely 
determine the steady-state distributions of species richness (as well as 
relative species abundances on local and large geographic scales). These two 
parameters are the so-called biodiversity number and the immigration 
(dispersal) rate. Under Hubbell's assumptions, the ecological properties of 
every individual in the population are assumed to be identical. 

In a neutral model of this type, individuals compete for the same pool of 
resources, but chance events are responsible for the identity of the final 
winner(s). The dynamics of each species is thus path-dependent and a Markovian 
description of their time evolution is appropriate. Under the assumption 
of a balance between birth, death and immigration rates, the neutral theory 
is able to reproduce the quantitative patterns of species distributions that 
are well known from the ecological literature. It also permits the generation 
of several nontrivial and testable quantitative predictions about biodiversity 
and biogeography. In particular, the theory predicts that rare species would 
typically be recent in terms of their origination. In relation to conservation 
biology, a neutral community in which species are essentially equal would be 
very fluid, with frequent replacements. If true, protected areas should be 
larger than those expected for stable communities with species closely 
adapted to given niches. 

Formally, Hubbell's theory is the ecological analog to the neutral theory 
of genetic drift in genetics (Kimura, 1983; Ewens, 1972; Karlin and McGregor, 
1972). Early attempts to incorporate the neutral approach from population 
genetics (Caswell, 1976; Hubbell, 1979) mainly highlighted the relevance of 
drift in community dynamics, providing evidence for a global view of 
ecosystems in which competitive forces, ecological niches, and even trophic 
interactions could be ignored in the pursuit of a better understanding of 
biodiversity dynamics. More recent work incorporated these ideas in an 
explicit way (Hubbell, 1997; Sol\'e and Alonso, 1998) and Hubbell's recent 
book provides an extensive, unifying account of these (Hubbell, 2001). The 
starting point of neutral models is a random community that evolves towards 
an equilibrium state under a stochastic birth-and death process incorporating 
dispersal. At high immigration rates, Hubbell's theory predicts a logseries 
distribution for the abundance of species in the local community, while when 
the immigration coupling between the metacommunity and the local community 
is lower, a lognormal-like shape is obtained for this distribution. Within
Hubbell's approximation, these distributions are shown to be particular 
cases of what he denotes as the zero-sum multinomial (Hubbell, 2001). 

Hubbell's model for local communities is similar to that proposed in Sol\'e 
{\em et al}., (2000) and analysed in McKane {\em et al}, 2000. There we took 
advantage of a mean field argument to find an analytical form for the 
stationary distribution for the probability of finding species having an 
abundance of $n$ individuals. In addition, we studied in detail its time 
behaviour using different approximations. Furthermore, our simplified 
approach based on this mean field argument allowed us to recover the scaling 
relationship between the fraction of links actualised and the number of 
species in a community --- the so-called $C^{*}$-$S$ relation, and gave
conditions in which such a relation arose.

Within Hubbell's mathematical framework the dynamical stochastic models 
were numerically solved and the equilibrium properties analysed. In this paper 
we present an analytic, general solution of Hubbell's model for the local 
community dynamics, that provides the stationary species-abundance 
distributions together with the time evolution from the initial state 
towards the stationary distribution.

\section{Formulation of the theory}
\label{formulation}

Hubbell's theory concerns populations on two scales: local communities and
regional metacommunities. To explain the model and derive the equations
in the simplest possible way, we will use the language of urn models
(Feller, 1968; Johnson and Kotz, 1977). This is a natural description when 
the stochastic dynamics in one time step only depends on the state of the 
system at the beginning of the time step (in other words is a Markov process). 
It also provides us with a concrete picture of the process which aids the 
derivation of the governing equation for the model. 

We begin by considering the model in a limit where the two levels of 
description are uncoupled. This allows us to focus only on the local 
community. We assume that there are $N_{i}$ individuals of species $i$ in the 
local community, with the total number of individuals of all species being $J$,
that is, $J = \sum^{r}_{i=1} N_{i}$ where $r$ is the total number of species.
The model is defined by picking one individual at random from the local
community, killing it, and then replacing it by an individual also drawn 
from the local community. In terms of the associated model this corresponds 
to having $N_i$ balls of colour $i$ ($i=1,\ldots,r$) in the urn. If we focus 
on one particular colour, $j$, the probability that the number of balls will 
decrease from $N_j$ to $N_{j} - 1$ during one time step is
\begin{equation}
W \left( N_{j} - 1 | N_{j} \right) = \frac{N_j}{J}\,
\frac{\left( (J-1) - (N_{j}-1) \right)}{J - 1}\,,
\label{down_noreg}
\end{equation}
since a ball of colour $j$ must be discarded and one of any other colour 
replaced for such a transition to occur. On the other hand, the probability 
that the number of balls will increase from $N_j$ to $N_{j} + 1$ requires 
that a ball of any other colour but $j$ must be discarded, and one of colour 
$j$ be replaced. Therefore 
\begin{equation}
W \left( N_{j} + 1 | N_{j} \right) = \frac{(J - N_j)}{J}\,
\frac{N_{j}}{J - 1}\,.
\label{up_noreg}
\end{equation}

The whole point of the model, however, is to couple local communities and 
regional metacommunities. This is achieved by choosing a replacement ball from 
the urn only $(1-m)$ of the time. For the rest of the time it is chosen from 
outside the urn. The probability of picking a ball of colour $j$ from this 
external source is defined to be $P_j$, and corresponds to assuming that the 
replacement individual comes from the regional metacommunity where species $i$
has a relative abundance of $P_i$. The transition probabilities 
(\ref{down_noreg}) and (\ref{up_noreg}) now read
\begin{equation}
W \left( N_{j} - 1 | N_{j} \right) = (1 - m) \frac{N_j}{J}\,
\frac{(J - N_{j})}{J - 1}\ + m \frac{N_j}{J}\,(1 - P_{j})
\label{down_full}
\end{equation}
and
\begin{equation}
W \left( N_{j} + 1 | N_{j} \right) = (1 - m) \frac{(J - N_j)}{J}\,
\frac{N_{j}}{J - 1} + m \frac{(J - N_{j})}{J} P_{j}\,.
\label{up_full}
\end{equation}

The change in the probability that there are $N_{j}$ balls in the urn from
time $t$ to the time after one time step has elapsed consists of four 
contributions. Two of these correspond to an increase in this probability
(due to transitions from $(N_{j}-1)$ and $(N_{j}+1)$ to $N_j$) and two to a
decrease (due to transitions from $N_{j}$ to $(N_{j}+1)$ and $(N_{j}-1)$). 
The balance equation showing this change is:
\begin{eqnarray}
\Delta P(N_{j}, t) &=& W(N_{j} | N_{j} - 1) P(N_{j} - 1, t)
+ W(N_{j} | N_{j} + 1) P(N_{j} + 1, t) \nonumber \\ \nonumber \\
&-& \left\{ W(N_{j} + 1 | N_{j}) 
+ W(N_{j} - 1 | N_{j}) \right\} P(N_{j}, t)\,. 
\label{discrete_mas}
\end{eqnarray}
Compared with the long time scales we are interested in --- during which 
many transitions will take place --- the step size is very small, and we may 
take the limit in which $\Delta P(N_{j}, t) \to dP(N_{j}, t)/dt$. The 
resulting equation is a master equation for the probability $P(N_{j}, t)$
(Van Kampen, 1981; Gardiner, 1985). Some care is needed with the boundary 
conditions on this equation: clearly the cases $N_{j} = 0$ and $N_{j} = J$ 
are special cases since there can be no transitions which reduce $N_{j}$ in 
the former case or which increase $N_j$ in the latter case. One possibility 
is to write two separate equations for these special cases. However there is
no need for this if we first observe that some of these conditions are 
natural consequences of the form of the transition probabilities. For 
example, the expressions in (\ref{down_full}) and (\ref{up_full}) are both
zero if $N_{j} = 0$ and $N_{j} = J$ respectively. So as long as we agree to
impose the formal definitions $W( 0 | -1 ) = 0$ and $W( J | J+1 ) = 0$ the 
same master equation may be used for all states. In addition, an initial
condition needs to be imposed to complete the specification of the problem. 
Typically, the number of individuals in the local community at $t=0$ will be
given: $P(N_{j}, 0) = \delta_{N_{j},N_{j,0}}$.

The mathematical formulation of Hubbell's theory described above can be 
directly mapped on to another dynamical model of a multispecies community
which we introduced a few years ago (Sol\'e {\em et al}., 2000; 
McKane {\em et al}., 2000; Sol\'e {\em et al}., 2002). In this case though, 
the nature of the interaction depends on the ``score'' between one species
and another, and a form of mean field theory had to be used in order to 
describe the dynamics by such a straightforward dynamics. In terms of the
notation we have used above --- $N$ denoting the number of individuals of a 
particular species and $J$ denoting the total number of individuals of all 
species --- the transition probabilities of this model are
(Sol\'e {\em et al}., 2000; McKane {\em et al}., 2000):
\begin{equation}
W( N + 1 | N ) = C^* (1-\mu) \frac{N}{J}\,\left( \frac{J - N}{J - 1} \right) 
+ {\mu \over S} \frac{J - N}{J}\,,
\label{gn}
\end{equation}
and
\begin{equation}
W( N - 1 | N ) = C^* (1-\mu) \frac{N}{J}\,\left( \frac{J - N}{J - 1} \right) 
+ {\mu \over S} (S-1) \frac{N}{J}\,. 
\label{rn}
\end{equation}
Here $\mu$ is the fraction of the time that replacing of one species by 
another can happen by chance, and not because the replacement individual 
belongs to a species which has a positive score against the first. It clearly
maps into $m$. The other constants are $S$, the number of species, and 
$C^{*}$, a parameter related to the degree of connectivity of the matrix of 
scores between the species. The precise form of the mapping is $C^{*} = 1$
and $P_{j} = S^{-1}$. 

Since we have analysed this model extensively (McKane {\em et al}., 2000) we
may simply deduce expressions for quantities of interest in the Hubbell 
theory by setting $C^{*} = 1, S = P_{j}^{-1}$ and $\mu = m$.

\section{Stationary state}
\label{stationary}

The most straightforward questions we can investigate concern the nature of
the stationary state of the theory. Let us begin by introducing the 
abbreviations
\begin{equation}
r_{N_j} \equiv W \left( N_{j} - 1 | N_{j} \right) = \frac{N_j}{J}\,
\left[ (1 - m) \frac{(J - N_{j})}{J - 1}\ + m (1 - P_{j}) \right]
\label{rN}
\end{equation}
and
\begin{equation}
g_{N_j} \equiv W \left( N_{j} + 1 | N_{j} \right) = \frac{(J - N_j)}{J}\,
\left[ (1 - m) \frac{N_{j}}{J - 1} + m P_{j} \right]\,.
\label{gN}
\end{equation}
The master equation now reads
\begin{equation}
\frac{dP(N_{j}, t)}{dt} = r_{N_{j}+1}\,P(N_{j} + 1, t) + 
g_{N_{j}-1}\,P(N_{j} - 1, t) - 
\left\{ r_{N_{j}} + g_{N_{j}} \right\} P(N_{j}, t) \,.
\label{master}
\end{equation}
The stationary probability distribution, $P_s(N_{j})$, is determined by 
setting $dP(N_{j})/dt=0$. This gives
\begin{equation}
r_{N_{j}+1}\,P_s(N_{j}+1) - g_{N_{j}}\,P_s(N_{j}) = r_{N_j}\,P_s(N_{j}) 
- g_{N_{j}-1}\,P_s(N_{j}-1)\,. 
\label{current}
\end{equation}
This is true for all $N_j$, which implies that 
$r_{N_{j}}\,P_s(N_{j}) - g_{N_{j}-1}\,P_s(N_{j}-1) = I$, where $I$ is a 
constant. Applying the boundary condition at $N_{j} = 0$, we find that $I = 0$ 
and therefore
\begin{equation}
r_{N_{j}+1}\,P_s(N_{j} + 1) = g_{N_{j}}\,P_s(N_{j})\ ; \ N_{j} = 0,1,...,J\,. 
\label{balance}
\end{equation}
To solve this equation, let us first assume that $m \neq 0$. Then the
$r_{N_j}$ and $g_{N_j}$ given by (\ref{rN}) and (\ref{gN}) are all non-zero
and we can solve (\ref{balance}) by iteration to obtain
\begin{equation}
P_s(N_{j}) = \frac{g_{N_{j}-1}\, g_{N_{j}-2} ...\, g_{0}}{r_{N_{j}}\, 
r_{N_{j}-1}\ldots\, r_{1}} P_s(0)\ ; 
\ N_{j} = 1,...,J\, .
\label{Ps}
\end{equation}
The constant $P_s(0)$ can be determined from the normalisation condition
\begin{equation}
\sum_{N_{j}=0}^J P_s(N_{j}) = P_s(0) + \sum_{N_{j}>0} P_s(N_{j}) = 1\,.
\label{normalization}
\end{equation}

To simplify the algebra let us introduce some new notation for various 
combinations of parameters which naturally appear in the solution of the model.
We write the transition probabilities as
\begin{equation}
r_{N_j} = \frac{(1 - m)}{J(J - 1)}\,N_{j}\,(N^{*}_{j} - N_{j} )\,,
\label{new_rN}
\end{equation}
and
\begin{equation}
g_{N_j} = \frac{(1 - m)}{J(J - 1)}\,(J - N_j) (N_{j} + P_{j}^{*})\,,
\label{new_gN}
\end{equation}
where
\begin{equation}
P_{j}^{*} = \frac{m(J- 1)}{(1 - m)}\,P_{j} \ \ {\rm and} \ \ 
N_{j}^{*} = \left( \frac{J - m}{1 - m} \right) - P_{j}^{*} \,.
\label{star_def}
\end{equation}
Substituting the expressions (\ref{new_rN}) and (\ref{new_gN}) into (\ref{Ps})
gives an explicit representation for the $P_{s}(N_{j})$ in terms of $P_{s}(0)$.
An expression for $P_{s}(0)$ itself can be obtained by performing the finite 
sum which appears in (\ref{normalization}). This sum can be performed 
analytically using properties of Jacobi polynomials (Abramowitz and Stegun, 
1965). Alternatively, the mapping into the model defined by (\ref{gn}) and 
(\ref{rn}) can be used since the result for the $P_{s}$ is known in this case
(Sol\'e {\em et al}., 2000; McKane {\em et al}., 2000). One finds (see 
McKane {\em et al}., 2000, for details of the derivation):
\begin{equation}
P_s(N_{j}) = {J \choose N_{j}} {\beta(N_{j}+P_{j}^{*}, N_{j}^{*}-N_{j}) 
\over \beta(P_{j}^{*}, N_{j}^{*}-J) }\,,  
\label{Ps_beta}
\end{equation}
where $\beta(a,b)=\Gamma(a) \Gamma(b) / \Gamma(a+b)$ is the beta-function. 

It is interesting to note that in the case $m=0$, where the local community 
is decoupled from the regional metacommunity, $g_{0}=0$, and so from 
(\ref{balance}), since $r_{1} \neq 0$, it follows that $P_{s}(1)=0$. In fact, 
since $r_{N_{j}} \neq 0$ for $0 < N_{j} < J$, we see from (\ref{balance}) that
$P_{s}(N_{j}) = 0$ for all $0 < N_{j} < J$. So with no interaction with the
regional metacommunity, species $j$ either disappears or becomes the only
species there is in the local community. Therefore some degree of coupling is
vital for biodiversity.

In Fig 1, we have computed the stationary distribution for different 
parameter values and sizes of the system. The relative species abundance 
distribution predicted to occur in local communities --- the zero-sum 
multinomial--- by the unified theory of Hubbell can be readily computed 
even for high community sizes using the analytic formula (\ref{Ps_beta}).  

\bigskip

\section{Time dependence}
\label{temporal}
Together with universal features displayed by the stationary 
patterns observed in mature communities, some common features 
are also observed when looking at how diversity develops in time. 
When an empty field starts to be colonized by
immigrant species a new community gets formed and a pattern of species
replacement develops. The transition from abandoned field to mature forest
is one of the best known examples of ecological sucession and is 
common in many places after the abandonment of agricultural
land. In temperate climates, a mature forest is the end point 
of sucession, in spite of the diferent potential initial 
conditions. The path towards the steady species-ranks distribution 
seems to be common to many different ecosystems (Hubbell, 2001). 
Furthermore, natural systems are continuously perturbed; any disturbance 
resumes the process of ecological succession. It is thus natural to ask: 
what predictions about this process can be made in the context of Hubbell's 
neutral theory? 

In the last section it was shown that a closed form expression could be 
obtained for the probability of finding $N_j$ individuals of species $j$ in 
the local community when the systems has reached the stationary state. In
addition to this, just mentioned, we also wish to know how the community 
is assembled from a given starting point. This requires us to solve for the 
time-dependence of the model. It is not possible, in general, to carry this 
out exactly, since the transition probabilities (\ref{new_rN}) and 
(\ref{new_gN}) are nonlinear functions of $N_j$. It is nevertheless possible 
to get a very good approximation to $P(N_{j}, t)$ by using the fact that in 
cases of interest $J$ will be large. The approach which we will use, due to 
Van Kampen (1981), is rather technical and has been discussed elsewhere in 
some detail (Van Kampen, 1981; McKane {\em et al}., 2000), but the basic idea 
is quite simple. Therefore, we will avoid these complications, and quote 
relevant results using the correspondence with the transition probabilities 
(\ref{gn}) and (\ref{rn}). 

The key idea is to expand about the deterministic version of the theory. In the
limit where the number of individuals becomes infinite, all stochasticity is 
lost, and the system is completely described by a deterministic equation. This
equation is not known a priori, but if it can be established, an expansion 
in powers of $J^{-1}$ could perhaps be set up to calculate corrections to the 
deterministic result which would be valid for large, but finite, $J$. 
Quite generally we would expect a plot of $P(N_{j}, t)$ against 
$N_{j}$ for fixed $t$ to be approximately Gaussian for large $J$. The motion 
of the peak of this distribution would move with $t$ according to the 
deterministic equation. Van Kampen's large $J$ expansion gives the 
deterministic equation as the zeroth order ($J \to \infty$) result, with the 
next to leading order result giving a Gaussian distribution peaked at this 
value. Higher order contributions give corrections to this distribution, but 
they are usually so small for large $J$ that they are of very little interest.
Since a Gaussian centred on a given value is completely determined by its 
width, there are only two things to find: (i) the deterministic equation, 
(ii) the width of the distribution.  

In practice one writes $N_{j} = J \phi_{j} (t) + J^{1/2} x_{j}$, where 
$\phi_{j} (t) = \lim_{J \to \infty} (N_{j}/J)$ is the fraction of $j$ species
which are present in the local community at time $t$ in the deterministic 
limit. The variable
\begin{displaymath}
x_{j} = \frac{1}{\sqrt{J}}\,\left( N_{j} - J \phi_{j} (t) \right)
\end{displaymath}
characterises the fluctuations away from the deterministic theory. We require 
$\phi_{j} (t)$ and $\langle x_{j}^{2} \rangle$ ($\langle x_{j} \rangle = 0$).
Using the correspondence between the two models we obtain 
(McKane {\em et al}., 2000) 
\begin{equation}
\frac{d\phi_{j}}{d\tau} = m \left( P_{j} - \phi_{j} \right)\,,
\label{macroscopic}
\end{equation}
where $\tau = t/J$ is a rescaled time. This equation is easily understood: if 
$\phi_{j}$ is less than the abundance of species $j$ in the regional 
metacommunity, then it increases. If it is more, then it decreases. The 
equation is easily solved to give
\begin{equation}
\phi_{j}(\tau) = \phi_{j}(0)e^{- m \tau} + P_{j}\,(1 - e^{- m\tau})\,.
\label{phi_tau}
\end{equation}
Initially we ask that $x_{j}(0) = 0$, which means that 
$\phi_{j}(0) = N_{j}(0)/J = N_{j,0}/J$. Going back to the $t$ variable gives
\begin{equation}
\phi_{j}(t) = \frac{N_{j,0}}{J}\, e^{- m t/J} + P_{j}\,(1 - e^{- m t/J})\,.
\label{phi_t}
\end{equation}

In Hubbell (2001, Chapter 4), an alternative discrete-time formulation of 
this local community model is given. Obviously, both time discrete and time 
continuous formulations give rise to the same equations for the deterministic 
model counterpart (Hubbell, 2001, page 110). However, he does not address 
the stochastic time-continuous formulation. Here we show that insight can 
be gained by finding approximate solutions to the time-dependent model.
  
The width of the distribution is given by
\begin{eqnarray}
\langle x_{j}^{2} \rangle_{\tau} & = & \frac{1}{m}\, 
P_{j}(1-P_{j})\, \left[ 1 - e^{-2 m \tau} \right] \nonumber \\
& + & {\cal A}_{j} \frac{2-m}{m}\,(1-2P_{j})\, e^{- m \tau}
\left[ 1 - e^{- m \tau} \right] - 2 (1 - m){\cal A}^{2}_{j}
\tau e^{-2 m \tau}\,, 
\label{var}
\end{eqnarray}
where ${\cal A}_{j} = (N_{j,0}/J) - P_{j}$. We have already commented that 
the probability distribution is a Gaussian to the order we have been working. 
Specifically, in terms of the quantities calculated above,
\begin{equation}
P(N_{j},t)=\frac{1}{\sqrt{2\pi J\, \langle x_{j}^{2}\rangle _{\tau}}}
\exp\left( -\frac{(N_{j}-J\, \phi_{j} (t))^{2}}{2 J\,  
\langle x_{j}^{2}\rangle _{\tau}}\right)\,, 
\label{Large-N}
\end{equation}
where $\phi_{j} (t)$ and $\langle x_{j}^{2}\rangle _{\tau}$ are given by 
equations (\ref{phi_t}) and (\ref{var}) respectively. 

In Fig. 2, we show the temporal evolution for $P(N_{j},t)$ computed both using
a Gaussian approximation (Eq. (\ref{Large-N})) and the numerical integration 
of the master equation. The good agreement which is obtained is a reflection 
of the fact that community sizes $J$ are taken to be large enough so that
further terms in the large $J$-expansion are negligible. However, if the 
final stationary distribution does not have a Gaussian shape, more terms 
should be included in the expansion so as to capture the true temporal 
behaviour of $P(N_{j},t)$. Notice that, while the approximation given by Eq. 
(\ref{Large-N}) is always represented as dotted or punctuated curves, in 
some cases these are not visible because they match the exact distribution
so completely. 

\section{Conclusion}
\label{conclude}
The main aim of this paper has been to show that aspects of Hubbell's neutral
model of local community biodiversity dynamics can be solved for exactly, and
even if this is not possible, calculational schemes are available which provide
very good approximations to the solution. Specifically, we have shown that the 
stationary properties of the model, which can be obtained from the zero-sum
multinomial, can all be found exactly. So, for instance, the mean value 
and variance of the number of individuals of species $j$, can be obtained from
this probability distribution. The nature of the time evolution cannot be 
determined in closed form, but a controlled approximation based on assuming 
that the total number of individuals of all species, $J$, is large, is 
possible. This is an excellent approximation in most cases of interest, and 
we would expect that the results that we have obtained will be relevant in 
these situations. The applicability of our approximation scheme was checked 
by carrying out the numerical integration of the master equation 
(Eq. \ref{master}). The results, displayed in Fig. 2, confirm our 
expectations. 

While the results which we have reported describe the essential aspects of 
the solution of Hubbell's model, there are many other interesting features 
which are also amenable to analysis and for which definite, and 
well-controlled, results may be obtained. The structure of the metacommunity 
and the form of the colonisation curve are examples. These, and related 
questions, are presently under study, and we hope to report our results in 
a future publication.

\vspace{0.7cm}

\section*{Acknowledgements}
DA would like to thank the MACSIN research group at the UFMG, Belo Horizonte,
Brazil for providing constant support and a nice working environment.
This work has been supported by a grant CIRIT FI00524 (DA) from the Catalan 
Government and by the Santa Fe Institute.
   
\newpage

\section*{References} 
Abramowitz, M. and Stegun, I.~A., 1965.
\newblock Handbook of mathematical functions.
\newblock Dover, New York. 

\medskip \noindent Bell, G., 2001.
Neutral macroecology. Science {\bf 293}, 2413-2418.

\medskip \noindent Brown, J.H., 1995.
Macroecology. The University of Chicago Press, Chicago.

\medskip \noindent Caswell, H., 1976. Community structure: a neutral model 
analysis. Ecol. Monogr. {\bf 46}, 327-354.

\medskip \noindent Ewens, W.~J., 1972. The sampling theory of selectively 
neutral alleles. Theor. Popul. Biol. {\bf 3}, 87-112.

\medskip \noindent Karlin, S. and McGregor, J., 1972. Addendum to a paper 
of W. Ewens. Theor. Popul. Biol. {\bf 3}, 113-116.

\medskip \noindent Kimura, M., 1983.
\newblock The neutral theory of molecular evolution.  
\newblock Cambridge University Press, Cambridge. 

\medskip \noindent Feller, W., 1968.
\newblock An introduction to probability theory and its applications.
\newblock Volume 1, Third edition. Wiley, New York.

\medskip \noindent Gardiner, C.~W., 1985. 
\newblock Handbook of stochastic methods.
\newblock Springer, Berlin. 2nd ed.

\medskip \noindent Hubbell, S.~P., 1997. 
A unified theory of biogeography and relative species abundance and 
its application to tropical rain forests and coral reefs.
Coral Reefs {\bf 16} (Suppl.), S9--S21.

\medskip \noindent Hubbell, S.~P., 2001. 
The unified theory of biogeography and biogeography.
Princeton University Press. Princeton, NJ.

\medskip \noindent Johnson, N.~L. and Kotz, S., 1977.
\newblock Urn models and their applications.
\newblock Wiley, New York.

\medskip \noindent Sol\'e, R. and Alonso, D., 1998.
Random walks, fractals and the origins of rainforest diversity. 
Adv. Complex Syst. {\bf 1}, 203-220.

\medskip \noindent McKane, A.~J., Alonso, D. and Sol\'e, R., 2000.
A mean field stochastic theory for species-rich assembled communities.
Phys. Rev. E {\bf 62}, 8466--8484.

\medskip \noindent Rosenzweig, M.~L., 1995. 
Species diversity in space and time. 
Cambridge University Press, Cambridge, UK.

\medskip \noindent Sol\'e, R., Alonso, D. and McKane, A.~J., 2000.
Scaling in a network model of a multispecies ecosystem.
Physica A {\bf 286}, 337--344.

\medskip \noindent Sol\'e, R., Alonso, D. and McKane, A.~J., 2002.
Self-organized instability in complex ecosystems.
Phil. Trans. R. Soc. Lond. B {\bf 357}, 667--681.

\medskip \noindent Van Kampen, N.~G., 1981. 
Stochastic processes in physics and chemistry. 
Elsevier, Amsterdam.
 
\medskip \noindent Wilson, E.~O., 2003. The encyclopedia of life. 
Trends in Ecology and Evolution {\bf 18}, 77-80. 

\newpage

\section*{Figure captions}

\begin{enumerate}

\item
Zero-sum multinomial distribution. The analytic formula (\ref{Ps_beta}) has 
been used to compute the stationary distribution, $P_{s} (N_{j})$, for 
different values of the abundance of species $j$ in the metacommunity, the 
total number of individuals $J$ and the probability of immigration from 
the metacommunity, $m$. We have dropped the subscript $j$, which labels a 
particular species, in the figure.  

\item
Temporal evolution of the probability, $P(N_{j}, t)$, of having the $j$-th 
species represented by $N_{j}$ individuals. The temporal evolution has been 
computed using both the Gaussian approximation and the straightforward 
numerical integration of the exact master equation. In both cases, the 
initial number of individuals of the focus species was $0.8 \times J$. The 
relative abundance of the focus species in the metacommunity was 
$P_{j} = 0.1$ also in both cases. We have dropped the subscript $j$, which 
labels a particular species, in the figure.

\end{enumerate}

\newpage
\begin{figure}[htb]
\vspace{18 cm}
\includegraphics{hubbell_fig1.eps}
\end{figure}
\vspace{3cm}
\centerline{\large Figure 1}

\newpage
\begin{figure}[htb]
\vspace{14 cm}
\includegraphics{hubbell_fig2.eps}
\end{figure}
\vspace{3cm}
\centerline{\large Figure 2}

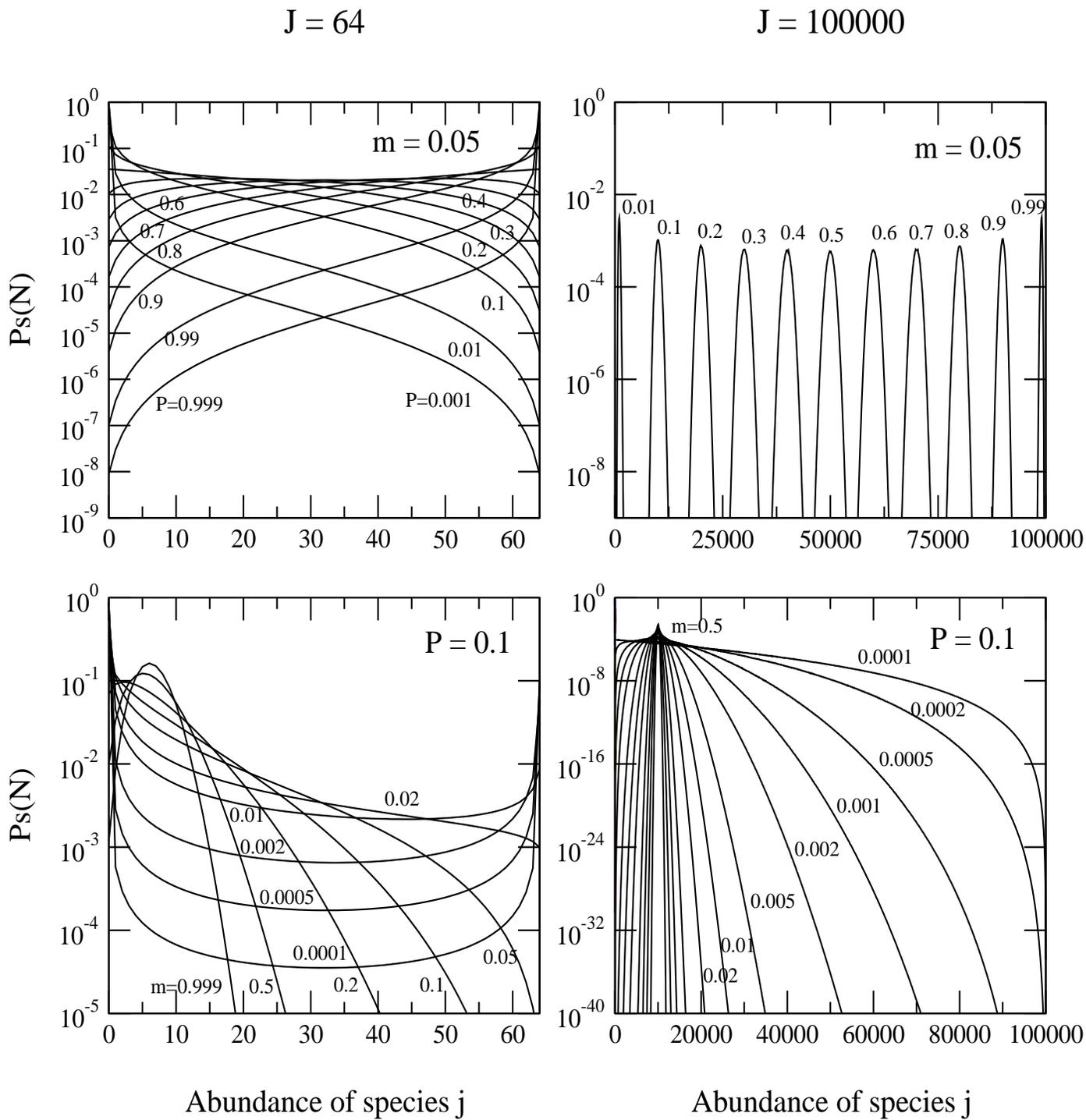
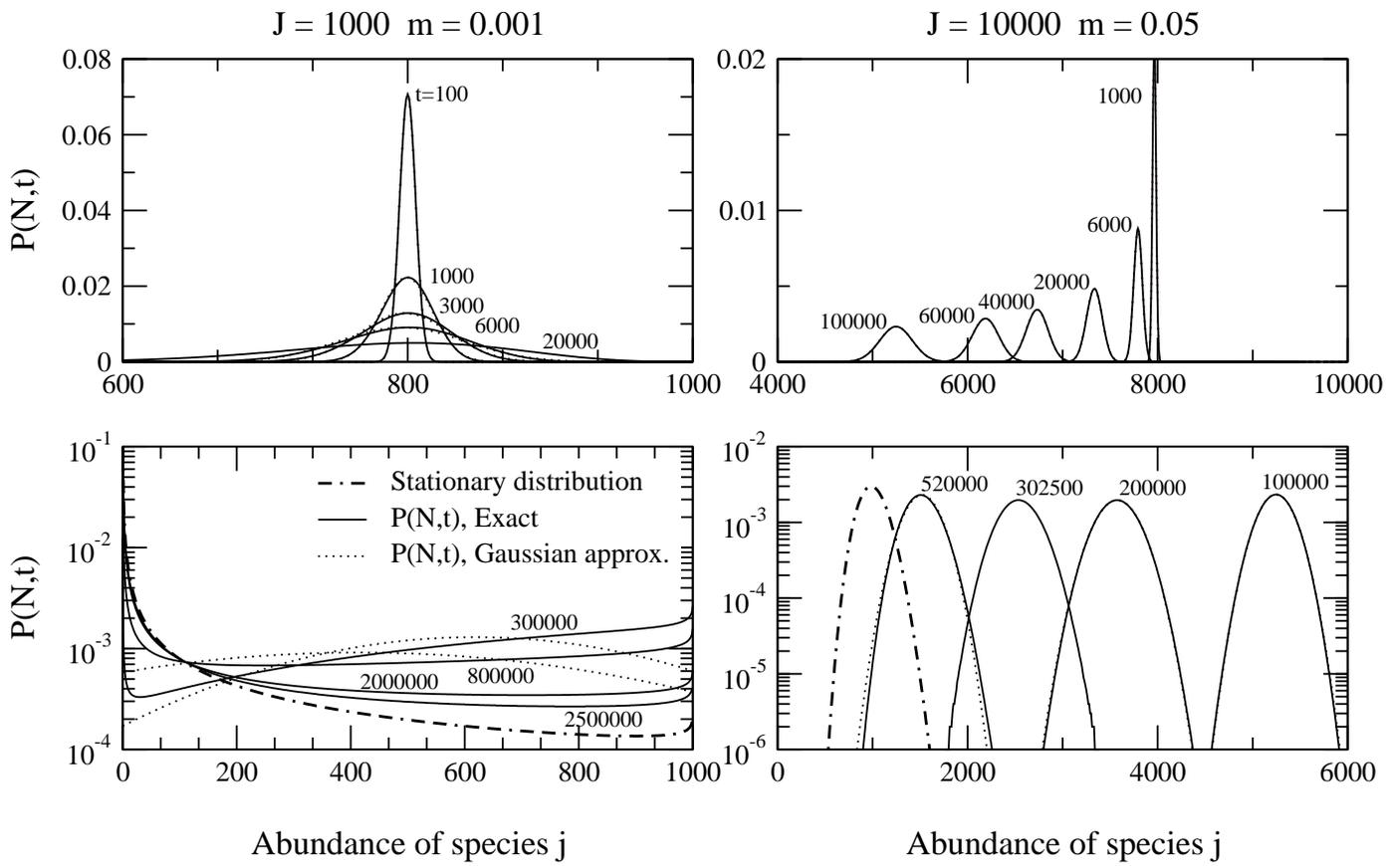
\end{document}